\documentclass[11pt,onecolumn]{article}
\usepackage[dvips]{graphics,color}
\usepackage{float}
\usepackage{graphicx}
\usepackage{epstopdf}
\usepackage{float}
\usepackage{latexsym,amssymb}
\usepackage{amsmath, amsbsy}
\usepackage{amsopn, amstext}
\usepackage{psfrag, epsfig}
\usepackage{cite}

\usepackage[noend]{algpseudocode}
\usepackage{algorithmicx,algorithm}
\usepackage[hidelinks]{hyperref}

\newtheorem{theorem}{Theorem}[section]
\newtheorem{definition}[theorem]{Definition}
\newtheorem{lemma}[theorem]{Lemma}

\newtheorem{example}[theorem]{Example}
\newtheorem{remark}[theorem]{Remark}

\newenvironment{proof}{\noindent\textit{Proof.}}

\makeatletter \@addtoreset{equation}{section}\makeatother

\newcommand{\be}{\begin{equation}}
\newcommand{\ee}{\end{equation}}

\newcommand{\bea}{\begin{eqnarray}}
\newcommand{\eea}{\end{eqnarray}}

\def\ba{\begin{array}}
\def\ea{\end{array}}

\renewcommand\baselinestretch{1.2}

\def\d{\delta}

\begin{document}
\begin{center}
{\Large\bf Minimum observability of probabilistic Boolean networks
 \footnote{This work is supported by National Natural Science Foundation of China under Grant 62103176, the ``Guangyue Young Scholar Innovation Team" of Liaocheng University under Grant LCUGYTD2022-01, and  the ``Discipline with  Strong Characteristics of Liaocheng University-Intelligent Science and Technology"  under Grant 319462208.\\
$\ast$ Corresponding author:\;Shihua\;Fu.\;Email:\;fush\_shanda@163.com.}} \vspace*{1.5\baselineskip}

{$\mathrm{Jiayi\ Xu}^1, \mathrm{Shihua\ Fu}^{1,\ast}, \mathrm{Liyuan\ Xia}^1, \mathrm{Jianjun\ Wang}^2$}

{\small
1. Research Center of Semi-tensor Product of Matrices: Theory and Applications, School of Mathematical Sciences, Liaocheng University, Liaocheng 252000, Shandong, P. R. China\\
2. School of Science and Technology, University of Camerino, Camerino 62032, Italy}
\end{center}

\begin{abstract}
 This paper studies the minimum observability of probabilistic Boolean networks (PBNs), the main objective of which is to add the fewest measurements to make an unobservable PBN become observable. First of all, the algebraic form of a PBN is established with the help of semi-tensor product (STP) of matrices. By combining the algebraic forms of two identical PBNs into a parallel system, a method to search the states that need to be $H$-distinguishable is proposed based on the robust set reachability technique. Secondly, a necessary and sufficient condition is given to find the minimum measurements such that a given set can be $H$-distinguishable. Moreover, by comparing the numbers of measurements for all the feasible $H$-distinguishable state sets, the least measurements that make the system observable are gained.
 Finally, an example is given to verify the validity of the obtained results.
\end{abstract}


\section{Introduction}
 \label{sec:introduction}
Boolean networks (BNs) are effective models for analyzing gene regulatory networks in systems biology, which was first proposed by Kauffman in 1969 \cite{MSEI}. In a BN, all nodes can take ``1" or ``0" in each discrete time, indicating that the gene is open or close. The evolution of each state variable is decided by the logical relationship among itself and its neighbor's states. Since the BNs are the simplest logical dynamical systems, the research of BNs has aroused the attention of many experts and scholars. At present, BNs have been widely used in systems biology such as cell differentiation \cite{HRAA}, immune response \cite{OLGR}, neural network \cite{DRWS} and so on. In addition to biological networks, BNs have been successfully applied in many other fields, including networked evolutionary game \cite{TEOS}, information security \cite{ANSM}, wireless sensor networks \cite{AMSF} and so on.

In order to depict the randomness and uncertainty of gene regulatory networks, Shmulevich et al. proposed probabilistic Boolean networks (PBNs) \cite{PBNA}. PBN is a random extension of deterministic BNs, and its state update follows a given probability distribution that controls the activation frequency of each subnetwork. PBN plays an important role both in theory and in the practical application of gene regulatory networks. A number of computational tools which facilitate the modelling and analysis of PBNs are developed, such as the mathematical theory of discrete-time Markov chains and integer programming approach. The dynamic behavior of PBNs can be modeled as Markov Chains. \cite{APCA} deals with the problem of optimal intervention for PBNs with the help of the theory of discrete-time Markov decision process. Integer programming can effectively solve the control problem of PBNs, such as optimal external control \cite{OBAT}.
Among theoretical topics, \cite{SAOG} studies steady-state analysis of PBNs by determining the number of iterations required to make the Markov chain converge. \cite{PAOP} solves the reachability problem of PBNs by a model checking approach.

Due to the lack of effective research tools, it is very difficult to systematically analyze the dynamic process of BNs and PBNs. The control problems of logical dynamic systems are studied even less. This situation was not solved until Cheng proposed the semi-tensor product (STP) of matrices \cite{ALRD}. Using the STP method, the logical form of BNs and PBNs can be transformed into equivalent algebraic forms, which is convenient for us to analyze and manipulate these logical networks according to the classical control theory. Up to now, many material problems of BNs and PBNs are studied and solved via STP, such as controllability \cite{C1,C2,C3,C4}, observability \cite{o1,o2,o5}, detectability \cite{d1,d2}, set stablization and set stability \cite{SSAS,li3,SFSF}, disturbance decoupling \cite{F1}, optimal control \cite{wu2,li1,li2} and so on. In addition to logical dynamic systems, STP has also been applied to many other fields, such as game theory \cite{z1,e1}, finite
automata \cite{S} and so on.

Observability is a fundamental and important concept in system science and control theory,  which elaborates a problem of whether the initial state can be identified  uniquely through the observed output sequences. In a BN, if any two different initial states can be distinguished by observing the output sequence, then the BN is observable. The existing results provide many methods to study the observability of BN, such as the graph-theoretic approach \cite{OOBN}, observability matrix-based analysis \cite{OOBM}, set controllability method \cite{OBNV,OBUP}, finite automata approach \cite{OOBC} and the weighted-pair-graph method \cite{EVOO}. Since the BN is a deterministic system, the state trajectory of each given initial state is unique. For a PBN, due to the uncertainty of the evolutionary dynamics, the output sequence of any given initial state can not be completely determined. Thus, compared with deterministic BNs, the observability of PBNs is more challenging. Several kinds of observability about PBNs have been provided \cite{OOPB,pin5}. In \cite{OOPB}, the definition of finite-time observability in probabilistic is given, which requires that for any two different initial states, the probability that their output sequences are not equal in finite time is
nonzero. Two different concepts of observability are defined in \cite{pin5}. One is finite-time observability with probability one, which requires that the probability of the corresponding output sequence being not equal in finite time is one for any two different initial states. The other case is asymptotic observability in distribution, which satisfies that for any two different initial states, when the time tends to infinity, the probability that the output is different tends to one. In addition,
\cite{OARO} proposes another representation of observability with probability one, and proposes a set theory algorithm to determine all possible initial states compatible with a given output.

When a given system cannot be observed by its original measurements, additional measurements are placed to make it observable. Considering the costs, the number of additional sensors required is crucial. How to find the minimum number of measurements to make the system observable, called the  minimum observability problem of the system, is a problem worth exploring.
In logical dynamic systems, the minimum observability problem of conjunctive BNs is first studied. Reference \cite{APAF} designs an effective algorithm to solve the minimum observability using graph theory approach. Subsequently, Liu. Y proposes a new method to study the minimum observability about general BNs, and obtains a method for determining the minimum states in which observations can be directly added \cite{o3}. Furthermore, \cite{o4} generalizes the result of minimum observability in \cite{o3} to the case with control inputs. Compared with
deterministic BNs, the stochastic model is more in line with real life.
However, as far as we know, there is no result about the minimum observability of PBNs.

In this paper, we investigate the minimum observability of PBNs for the first time. To solve the minimum observability of PBNs, we need to address two key issues: (i) determine which state sets
need to be $H$-distinguishable such that an unobservable PBN can become observable;
(ii) add the least measurements to make a give state set  $H$-distinguishable.
For an unobservable PBN, the distinguishable states are not affected by the newly added observations. Therefore, this paper mainly studies the dynamic trajectories of indistinguishable states. In \cite{o3,o4}, the minimum states that need to become initially distinguishable are determined by the attractors of the BNs and BCNs. However, since the state trajectories of PBNs are too complex, the method of \cite{o3,o4} is not suitable for PBNs. We study the minimum observability of PBNs mainly based on the robust reachability technique. The method in this paper is more general, and is also applicable to deterministic BNs. The main contributions of this paper are as follows.

(i) A method to make an unobservable PBN observable is proposed. A class of minimal state sets are obtained by the robust set reachability technique, as long as one of these sets becomes $H$-distinguishable, the other indistinguishable states can become distinguishable, and then the PBN becomes observable.

(ii) The minimum observability problem of PBN is solved. By constructing a group of truth value matrices, all the schemes (including the optimal scheme) that make the given sets $H$-distinguishable are obtained. By comparing the numbers of measurements for the feasible $H$-distinguishable state sets, the minimum measurements added to make the system observable is determined.

The rest of this article is organized as follows. Part 2 contains some commonly used symbols and preliminaries. Main results of this article are in Part 3. An illustrative example is given in Part 4. Some brief conclusions are provided in Part 5.

\section{Preliminaries}
\subsection{Symbols and STP of matrices}
We first give some commonly used symbols, as shown in Table 1.

\begin{table}[ht]
\centering
 \caption{Notations}
 \vspace{2.5mm}
\begin{tabular}{c c}
\hline
Notations & Definitions\\
\hline
$\mathcal{D}$ & $\{1,0\}$\\
$\mathcal{D}^n$ & $\underbrace{\mathcal{D}\times\cdots\times\mathcal{D}}_n$\\
$\mathbf{Z}_+$ &  set of positive integers\\
$\mathbb{R}_{m\times n}$ &  set of $m\times n$ real matrices\\
$\mathrm{Col}_i(M)$ & the $i$-th column of matrix $M$.\\
$\mathrm{Row}_i(M)$ & the $i$-th row of matrix $M$\\
$\delta_n^r$ & $\mathrm{Col}_r(I_n)$\\
$\Delta_n$ & $\{\delta_n^r \mid r=1,\cdots,n\}$\\
$\mathrm{Col}(M)\subseteq \Delta_m$ & $M\in\mathbb{R}_{m\times n}$ is a logical matrix\\
$\mathcal{L}_{m\times n}$ & set of $m\times n$ logical matrices\\
$L=\d_m[i_1\ i_2 \cdots i_n]$ & matrix $L$ with $\mathrm{Col}_s(L)=\d_m^{i_s}$\\
$L_{i, j}$ & the $(i,j)$-element of matrix $L$\\
$\mid \cdot\mid$ & the cardinality of a set\\
$[L]_{i,j}\in\{0,1\}$ & $L$ is a Boolean matrix.\\
$\mathcal{B}_{m\times n}$ & set of $m\times n$ Boolean matrices\\
$M+_\mathcal{B}N$ & $[M+_\mathcal{B}N]_{i,j}=[M]_{i,j}\vee[N]_{i,j}$\\
$L^\top$ &  the transpose of matrix $L$\\
$\mathbf{1}_n$ & $n$-dimensional column vector whose entries are equal to 1.\\
$lcm(a_1,a_2,\ldots,a_s)$ & the least common multiple of $a_1,a_2,\ldots,a_s$\\
$a\in\mathcal{A}\setminus\mathcal{B}$ & $a\in\mathcal{A},a\notin\mathcal{B}$\\
$\otimes$ & Kronecker product of matrices\\
$\ast$ & Khatri-Rao product of matrices\\
\hline
\end{tabular}
\end{table}

{\definition \cite{ALRD}\label{d2.2.1} The $\mathrm{STP}$ of $C\in \mathbb{R}_{a\times b}$ and $D\in \mathbb{R}_{q\times r}$ is defined as
$$C\ltimes D=(C\otimes I_{\frac{\lambda}{b}})(D\otimes I_{\frac{\lambda}{q}}),$$
where $\lambda=lcm(b,q)$.}

{\remark  The $\mathrm{STP}$ is a generalization of the traditional matrix multiplication, which preserves almost all the properties of traditional matrix multiplication, thus ``$\ltimes$'' can be omitted in the following.}
{\definition\hskip -0.1mm\cite{ZALJ}
The Khatri-Rao product of $M\in {\cal M}_{m\times r}$ and $N\in{\cal M}_{n\times r}$ is defined as
$$M\ast N=[\mathrm{Col}_1(M)\otimes \mathrm{Col}_1(N), \mathrm{Col}_2(M)\otimes \mathrm{Col}_2(N),\ldots, \mathrm{Col}_r(M)\otimes \mathrm{Col}_r(N)].$$}

Identify ``$1\sim\delta_2^1$" and ``$0\sim\delta_2^2$", where ``$\sim$" represents an equivalence relation, then $\Delta \sim \mathcal{D}$. We call $\delta_2^1$ and $\delta_2^2$ the vector form of $1$ and $0$, respectively. Based on the vector form of the logic variables, a Boolean function can be expressed as equivalent algebraic form.

\begin{lemma} \label{101}\cite{ALRD}
Let $ f:\mathcal{D}^r\rightarrow \mathcal{D}$ be a Boolean mapping, then there exists a unique $M_f\in\mathcal{L}_{2\times 2^r}$ such that
$$f(X_1,X_2,\ldots,X_r)= M_f\ltimes_{i=1}^rx_i,$$
where $x_i\in \Delta,\ i=1, 2, \cdots, r$ is the vector form of $X_i\in \mathcal{D}$, and $M_f$ is the structure matrix of $f$.
\end{lemma}
\subsection{Algebraic representation of PBNs}
A PBN with $n$ state nodes and $q$ output nodes is described as
\begin{eqnarray}\label{1}
\begin{cases}
x_i(t+1)=f_i^{\sigma(t)}(x_1(t),x_2(t),\ldots,x_n(t)),\ \  i=1,2,\ldots,n,&\\
y_j(t)=h_j(x_1(t),x_2(t),\ldots,x_n(t)),\ \  j=1,2,\ldots,q,&\\
\end{cases}
\end{eqnarray}
where $x_i(t)\in\mathcal{D}$ and $y_j(t)\in\mathcal{D}$ are the state and the output of the system (\ref{1}) at time $t$, respectively. $f_i^\sigma$, $h_j$ are logical functions. The switching signal $\sigma(t)$ is a stochastic sequence taking values from $[1:m]$, where $m$ is a positive integer denoting the number of subnetworks.

Using the vector form of variables and Lemma (\ref{101}), PBN (\ref{1}) can be represented by
\begin{eqnarray}\label{2}
\begin{cases}
x(t+1)=\mathbf{L}_{\sigma(t)}x(t),&\\
y(t)=Hx(t),&\\
\end{cases}
\end{eqnarray}
where $x(t)=\ltimes_{i=1}^nx_i(t)\in\Delta_{2^n}$, $y(t)\in\Delta_{2^q}$, and $\mathbf{L}_v\in\mathcal{L}_{2^n\times2^n}, v\in[1:m]$ is the structure matrix of the $v$-th sub-network. Identify $v\sim\delta_m^v$, where $v\in[1:m]$, then PBN (\ref{2}) can be expressed equivalently as
\begin{eqnarray}\label{3}
\begin{cases}
x(t+1)=\mathbf{L}\sigma(t)x(t),&\\
y(t)=Hx(t),&\\
\end{cases}
\end{eqnarray}
where $\sigma(t)\in\Delta_m$ and $\mathbf{L}=[\mathbf{L}_1\ \mathbf{L}_2\ \ldots\ \mathbf{L}_m]$.

Suppose that the stochastic switched signal $\sigma(t)$ is an independent and identically distributed process with a probability distribution of $\mathbb{P}\{\sigma(t)=\delta_m^i\}=p_i^\sigma, i\in[1:m]$, where $0\leq p_i^\sigma\leq1$ and $\sum_{i=1}^m p_i^\sigma=1$. Denote the probability distribution vector of $\sigma(t)$ as $\mathbf{p}^\sigma=[p_1^\sigma\ p_2^\sigma\ \ldots\ p_m^\sigma]^\top$, then the one-step state transition probability matrix of PBN (\ref{3}) can be expressed as
\begin{equation}\label{4}
\mathbf{P}^x=\sum_{i=1}^mp_i\mathbf{L}_i=\mathbf{L}\ltimes\mathbf{p}^\sigma,
\end{equation}
where $\mathbf{P}_{i,j}^x=\mathbb{P}\{x(t+1)=\delta_{2^n}^i|x(t)=\delta_{2^n}^j\}$. 

Let $\sigma^s$ represent a set of switching signal sequence with length $s$, $\sigma^s=(\sigma(0), \sigma(1), \ldots, \sigma(s-1))$,  and denote the set of all switching signal sequences with length $s$ as $\Sigma^s$. Denote the state of system (\ref{2}) at time $s$ starting from $x_0$ under switching signal $\sigma^s\in\Sigma^s$ as $x(s;\sigma^s,x_0)$, and denote the corresponding output as $y(s;\sigma^s,x_0)$, i.e., $y(s;\sigma^s,x_0)=Hx(s;\sigma^s,x_0)$.

\subsection{Observability of PBNs}
\begin{definition}\cite{pin5}
PBN (\ref{3}) is observable with probability one in finite time, if for any distinct initial state pair $x_0, \overline{x}_0\in\Delta_{2^n}$, there exists a positive integer $T$ such that
\begin{equation}\label{5}
\mathbb{P}\{(y_0,y(1;\sigma^1,x_0),\ldots,y(T;\sigma^T,x_0))\neq (y_0,y(1;\sigma^1,\overline{x}_0),\ldots,y(T;\sigma^T,\overline{x}_0))\}=1.
\end{equation}
These two states $x_0, \overline{x}_0$ are called distinguishable states. Otherwise, they are indistinguishable states.
\end{definition}
With the parallel extension method, the observability problem of PBN (\ref{3}) can be transformed into the set reachability problem of an augmented system. The main steps can be summarized as follows.

First of all, a duplicate of PBN (\ref{3}) is copied as
\begin{eqnarray}\label{6}
\begin{cases}
x'(t+1)=\mathbf{L}\sigma(t)x'(t),&\\
y'(t)=Hx'(t).
\end{cases}
\end{eqnarray}

Let $z(t):=x(t)\ltimes x'(t)$, $w(t):=y(t)\ltimes y'(t)$, an augmented system can be constructed as
\begin{align}  \label{7}\nonumber
z(t+1)= & x(t+1)\ltimes x'(t+1)\\ \nonumber
      = & (\mathbf{L}\sigma(t)x(t))\ltimes (\mathbf{L}\sigma(t)x'(t))\\ \nonumber
      = & [\mathbf{L}(I_{m2^n}\otimes\mathbf{1}_{2^n}^\top)\sigma(t)x(t)x'(t)]\ltimes [\mathbf{L}(I_m\otimes\mathbf{1}_{2^n}^\top)\sigma(t)x(t)x'(t)]\\ \nonumber
      = & \{[\mathbf{L}(I_{m2^n}\otimes\mathbf{1}_{2^n}^\top)]\ast[\mathbf{L}(I_m\otimes\mathbf{1}_{2^n}^\top)]\}\sigma(t)z(t)\\
      := & \mathbf{F}\sigma(t)z(t),\\ \nonumber
w(t)= & Hx(t)Hx'(t)\\ \nonumber
    = & H(I_{2^n}\otimes H)x(t)x'(t)\\ \nonumber
    = & (H\otimes I_{2^n})(I_{2^n}\otimes H)z(t)\\ \nonumber
    = & (H\otimes H)z(t).
\end{align}
Divide $\mathbf{F}$ into $m$ equal blocks according to the dimension of switching signal as $\mathbf{F}=[\mathbf{F}_1\ \mathbf{F}_2\ \ldots\ \mathbf{F}_m]$, where $\mathbf{F}_i\in\mathcal{L}_{{2^{2n}}\times {2^{2n}}}, i\in[1:m]$.
Let $\mathbf{Q}=\sum_{i=1}^mp_i^\sigma \mathbf{F}_i=\mathbf{F}\ltimes \mathbf{p}^\sigma$ and $K=H\otimes H\in \mathcal{L}_{{2^{2q}}\times {2^{2n}}}$, then the expectation of the augmented system (\ref{7}) can be expressed as
\begin{eqnarray}\label{8}
\begin{cases}
\mathbb{E}\{z(t+1)\}=\mathbf{Q}\mathbb{E}\{z(t)\},&\\
w(t)=Kz(t).
\end{cases}
\end{eqnarray}

Obviously, for any $z_0=x_0\ltimes x'_0$, we have
\begin{equation}\nonumber
z(t;\sigma^t,z_0)=x(t;\sigma^t,x_0)\ltimes x'(t;\sigma^t,x'_0),
\end{equation}
where $z(t;\sigma^t,z_0)$ represents the state starting from initial state $z_0$ of system (\ref{8}) at time $t\in\mathbb{Z}_+$ under switching sequence $\sigma^t\in\Sigma^t$.

\begin{definition}
Consider two states $x, x'\in\Delta_{2^n}$ and $z=x\ltimes x'\in\Delta_{2^{2n}}$. $z$ is named distinguishable, if $x, x'$ are distinguishable.
\end{definition}

For two state sets $\mathcal{Z}_1=\{\delta_{2^n}^i\delta_{2^n}^j\mid i<j\}$ and $\mathcal{Z}_2=\{\delta_{2^n}^j\delta_{2^n}^i\mid i<j\}$. If $\mathcal{Z}_1$ is a distinguishable set, then $\mathcal{Z}_2$ is distinguishable, otherwise, $\mathcal{Z}_2$ is indistinguishable, i.e., $\mathcal{Z}_1$ and $\mathcal{Z}_2$ have the same distinguishability. Therefore, in order to decrease computational complexity, we only decect the distinguishability of $\mathcal{Z}_1$.
Then three subsets are defined as
\begin{eqnarray}\label{9}
\begin{cases}
\mathcal{S}_0=\{\delta_{2^n}^i\delta_{2^n}^i\mid i=1,2,\ldots,2^n\},&\\
\mathcal{S}_1=\{\delta_{2^n}^i\delta_{2^n}^j\mid H\delta_{2^n}^i=H\delta_{2^n}^j, i<j\},&\\
\mathcal{S}_2=\{\delta_{2^n}^i\delta_{2^n}^j\mid H\delta_{2^n}^i\neq H\delta_{2^n}^j,i<j\},&\\
\end{cases}
\end{eqnarray}
where $\mathcal{S}_0$ is diagonal subset, $\mathcal{S}_1$ is $H$-indistinguishable subset, and $\mathcal{S}_2$ is $H$-distinguishable subset.

Subsequently, the observability of PBN (\ref{3}) is explored based on the set reachability of PBNs \cite{pin5}.

\begin{definition}\cite{pin5}
Suppose that $\mathcal{C}_0, \mathcal{C}_d\subset\Delta_{2^{2n}}$ are the initial and target sets, respectively. $\mathcal{C}_d$ is said to be finite-time reachable with probability one from $\mathcal{C}_0$ , if there exists a positive integer $k$ such that for any initial state $z_0\in \mathcal{C}_0$, it holds that
\begin{equation}\nonumber
\mathbb{P}\{z(k;\sigma^k,z_0)]\in \mathcal{C}_d\}=1
\end{equation}
for any $\sigma^k\in\Sigma^k$.
\end{definition}

\begin{lemma}\cite{pin5}\label{2.10}
PBN (\ref{3}) is observable with probability one in finite time, if and only if the set $\mathcal{S}_2$ is reachable from $\mathcal{S}_{1}$ with probability one in finite time in system (\ref{8}).
\end{lemma}

Lemma \ref{2.10} converts the observability of PBN (\ref{3}) into the set reachability problem of augmented system (\ref{8}). If there exists one state $\delta_{2^{2n}}^s\in\mathcal{S}_{1}$ such that $\delta_{2^{2n}}^s$ can not reach $\mathcal{S}_{2}$, then PBN (\ref{3}) is unobservable, and $\delta_{2^{2n}}^s$ is an indistinguishable state. In the next section, we consider to add  the least measurements to make the PBN (\ref{3}) observable.

\section{Main Results}

\subsection{The method to make a PBN observable}

In this subsection, we consider to solve the first issue, that is, search the sets that need to be $H$-distinguishable such that an unobservable PBN can become observable.

For PBN (\ref{3}), suppose that for any
$j_a\in\{1,2,\ldots,n\}$, $x_{j_a}$ can be directly measurable
under the new measurement $y_{q+a}=x_{j_a}$.
We study the observability of PBNs via STP. That is to say, a new measurement set $\mathcal{I}=\{x_{j_1}, x_{j_2}, \ldots , x_{j_m}\}$ is added to make the original unobservable system observable. Denote the added measurements as
\begin{eqnarray}
y_{q+1}(t)=x_{j_1}(t), \ldots, y_{q+m}(t)=x_{j_m}(t).
\end{eqnarray}
Let $\overline{y}(t)=y_{q+1}\ltimes y_{q+2}\ltimes\ldots\ltimes y_{q+m}$, the added outputs can be represented by $\overline{y}(t)=\overline{H}x(t)$. Further, new outputs can be depicted as
\begin{eqnarray}\label{3.14}
\begin{array}{l}
\begin{cases}
y_j(t)=h_j(x_1(t), x_2(t), \ldots, x_n(t)), j=1, \ldots, q,\\
y_{q+1}(t)=x_{j_1}(t),\\
y_{q+2}(t)=x_{j_2}(t),\\
\vdots\\
y_{q+m}(t)=x_{j_m}(t).
\end{cases}\\
\end{array}
\end{eqnarray}

Denote $y^\ast(t)=y_1\ltimes y_2\ltimes\ldots\ltimes y_{q+m}$, system (\ref{3.14}) can be represented by $y^\ast(t)=Gx(t)$, where $G=H*\overline{H}$.

Evidently, if $x, x'\in\Delta_{2^n}$ are distinguishable under original output, then $x, x'$ are distinguishable under the new output. Thus, we only need to make the indistinguishable states become distinguishable under new added measurements.
\begin{lemma}\label{33333}\cite{o4}
Suppose that $z=x\ltimes x'$ is indistinguishable under original PBN (\ref{3}), and $x=\ltimes_{r=1}^n\delta_{2^n}^{i_r}$, $x'=\ltimes_{r=1}^n\delta_{2^n}^{j_r}$. If $\delta_2^{i_k}\neq\delta_2^{j_k}$, then $x, x'$ ($z$) will become $H$-distinguishable under new added measurement $y(t) = x_k(t)$.
\end{lemma}

According to Lemma \ref{2.10}, for an unobservable PBN, there exists at least one indistinguishable state in $\mathcal{S}_{1}$ which can not reach $\mathcal{S}_{2}$. We need to find all these states and make them distinguishable. We first calculate the states in $\mathcal{S}_{1}$ that can reach $\mathcal{S}_{2}$ with probability one. We first give the following definition.

\begin{definition}
For a given set $\Lambda\subseteq\Delta_{2^{2n}}$, define its index vector $\mathbf{J}_\Lambda\in\mathcal{B}_{2^{2n}\times1}$ as follow:
\begin{equation}\nonumber
(\mathbf{J}_\Lambda)_i=
\begin{cases}
1,\ \delta_{2^{2n}}^i\in\Lambda,\\
0,\ \delta_{2^{2n}}^i\notin\Lambda.
\end{cases}
\end{equation}
\end{definition}

In the following, we calculate the reachable set at each step of $\mathcal{S}_{2}$ based on the vector representation of sets. Calculate a set
\begin{align}\label{3.1}
R_1(\mathcal{S}_{2})=\{\delta_{2^{2n}}^j|\mathbf{J}_{\mathcal{S}_{2}}^\top \mathbf{Q}\delta_{2^{2n}}^j=1, \delta_{2^{2n}}^j\notin\mathcal{S}_{2}\},
\end{align}
where $\mathbf{J}_{\mathcal{S}_{2}}^\top$ is the index vector of set $\mathcal{S}_{2}$. For any $\delta_{2^{2n}}^j\in R_1(\mathcal{S}_{2})$, it holds that
\begin{align}\nonumber
1
= & \mathbf{J}_{\mathcal{S}_{2}}^\top \mathbf{Q}\delta_{2^{2n}}^j\\ \nonumber
= & {\sum}_{\delta_{2^{2n}}^i\in\mathcal{S}_{2}}\mathbf{Q}_{i,j}\\ \nonumber
= & {\sum}_{\delta_{2^{2n}}^i\in\mathcal{S}_{2}}\mathbb{P}\{z(t+1)=\delta_{2^{2n}}^i|z(t)=\delta_{2^{2n}}^j\}\\ \nonumber
= & \mathbb{P}\{z(t+1)\in\mathcal{S}_{2}|z(t)=\delta_{2^{2n}}^j\}.
\end{align}
Thus $\delta_{2^{2n}}^j$ can reach $\mathcal{S}_{2}$ with probability one in one step. It is obvious that $R_1(\mathcal{S}_{2})$ is the one-step reachable set of $\mathcal{S}_{2}$ with probability one.

Similarly, we can calculate set
\begin{align}\nonumber
R_2(\mathcal{S}_{2})
=& R_1(\mathcal{S}_{2}\cup R_1(\mathcal{S}_{2}))\\ \nonumber
= & \{\delta_{2^{2n}}^j|(\mathbf{J}_{\mathcal{S}_{2}}^\top+_\mathcal{B}\mathbf{J}_{R_1(\mathcal{S}_{2})}^\top) \mathbf{Q}\delta_{2^{2n}}^j=1, \delta_{2^{2n}}^j\notin\mathcal{S}_{2}\cup R_1(\mathcal{S}_{2})\}. &\nonumber
\end{align}
For any $\delta_{2^{2n}}^j\in R_2(\mathcal{S}_{2})$, one has
\begin{align}\nonumber
1= & (\mathbf{J}_{\mathcal{S}_{2}}^\top+_\mathcal{B}\mathbf{J}_{R_1(\mathcal{S}_{2})}^\top) \mathbf{Q}\delta_{2^{2n}}^j\\ \nonumber
= & {\sum}_{\delta_{2^{2n}}^i\in\mathcal{S}_{2}\cup R_1(\mathcal{S}_{2})}\mathbf{Q}_{i,j}\\ \nonumber
= & {\sum}_{\delta_{2^{2n}}^i\in\mathcal{S}_{2}\cup R_1(\mathcal{S}_{2})}\mathbb{P}\{z(t+1)=\delta_{2^{2n}}^i|z(t)=\delta_{2^{2n}}^j\}\\ \nonumber
= & \mathbb{P}\{z(t+1)\in\mathcal{S}_{2}\cup R_1(\mathcal{S}_{2})|z(t)=\delta_{2^{2n}}^j\}\\ \nonumber
= & \mathbb{P}\{z(t+2)\in\mathcal{S}_{2}|z(t)=\delta_{2^{2n}}^j\}.
\end{align}
We can easily obtain that $R_2(\mathcal{S}_{2})$ is the two-step reachable set of $\mathcal{S}_{2}$ with probability one. Since $\delta_{2^{2n}}^j\notin\mathcal{S}_{2}\cup R_1(\mathcal{S}_{2})$, $\delta_{2^{2n}}^j$ can not reach $\mathcal{S}_{2}$ in one step with probability one.

Recursively, we calculate a set of sets
\begin{align}\label{3.2}
&R_k(\mathcal{S}_{2})
=R_1(\mathcal{S}_{2}\cup R_1(\mathcal{S}_{2})\cup\ldots\cup R_{k-1}(\mathcal{S}_{2}))\\ \nonumber
= &
\{\delta_{2^{2n}}^j|(_\mathcal{B}\sum_1^{k-1}\mathbf{J}_{R_i(\mathcal{S}_{2})}^\top+_\mathcal{B}\mathbf{J}_{\mathcal{S}_{2}}^\top)
\mathbf{Q}\delta_{2^{2n}}^j=1, \delta_{2^{2n}}^j\notin \mathcal{S}_{2}\cup R_1(\mathcal{S}_{2})\cup\ldots\cup R_{k-1}(\mathcal{S}_{2})\},\ where\ k>2.
\end{align}
Obviously,
$R_k(\mathcal{S}_{2})$ is the $k$-step reachable set of $\mathcal{S}_{2}$ with probability one.

Since the number of states of a PBN is finite, there must exist a positive integer $t^\ast\leq2^{2n}$ such that $R_{t^\ast}(\mathcal{S}_{2})\neq\emptyset, R_{t^\ast+1}(\mathcal{S}_{2})=\emptyset$. According to the above analysis, we can get Algorithm 1.
\begin{algorithm}[htp]
	\renewcommand{\algorithmicrequire}{\textbf{Require:}}
	\renewcommand{\algorithmicensure}{\textbf{Ensure:}}
	\caption{:Calculation of the reachable set of $\mathcal{S}_{2}$ with probability one.}
	\label{ppp1}
	\begin{algorithmic}[1]
		\Require $\mathbf{J}_{\mathcal{S}_{2}}, \mathbf{Q}$
\Ensure $\bigcup_{i=1}^{t^\ast}R_i(\mathcal{S}_{2})$
\State{\textbf{Initialize:} $k$;}
\State{Calculate $R_1(\mathcal{S}_{2})$ by (\ref{3.1});}
\State{\textbf{if}
$R_1(\mathcal{S}_{2})=\emptyset$ \textbf{stop}}
\State \textbf{else}
Go to 5;
\State {\textbf{while} $k\in[2:2^{2n}]$
\textbf{do}}
\State{Calculate $R_k(\mathcal{S}_{2})$ by (\ref{3.2});}
\State{\textbf{if}
$R_{k-1}(\mathcal{S}_{2})\neq\emptyset, R_k(\mathcal{S}_{2})=\emptyset$ \textbf{then}}
\State{$R_{t^\ast}(\mathcal{S}_{2})=R_k(\mathcal{S}_{2})$;}
\State \textbf{else}
$k=k+1$;
        \State Go to 5;
\State \textbf{end if}
\State \textbf{end while}
	\end{algorithmic}
\end{algorithm}

\begin{remark}
Algorithm \ref{ppp1} shows that the set of all states that can reach $\mathcal{S}_{2}$ with probability one is $\bigcup_{i=1}^{t^\ast}R_i(\mathcal{S}_{2})$. Denote $\mathcal{M}:=[\bigcup_{i=1}^{t^\ast}R_i(\mathcal{S}_{2})]\cap\mathcal{S}_{1}$, then $\mathcal{M}$ contains all the states in $\mathcal{S}_{1}$ which are distinguishable.
\end{remark}

 Let $\mathcal{N}=\mathcal{S}_{1}\backslash \mathcal{M}$, then $\mathcal{N}$ denotes the set of indistinguishable states in $\mathcal{S}_1$.
 To make the system become observable, appropriate measurements must be added to make the states in $\mathcal{N}$ distinguishable. Adding measurements directly to make all states in $\mathcal{N}$ be $H$-distinguishable is one method, but it may add a lot of redundant measurements. Since a measurement can make multiple states $H$-distinguishable, the measurements that need to be added can not be  determined just by the number of states.
  In the following, we search all the feasible minimal sets that need to be $H$-distinguishable. The system becomes observable only if all states in a minimal set are $H$-distinguishable.

 Define a state set in $\mathcal{N}$ as follow:
\begin{eqnarray}\label{3.13}
\mathcal{N}_1=\{\delta_{2^{2n}}^j|\mathbf{J}_{\mathcal{S}_0}^\top \mathbf{Q}\delta_{2^{2n}}^j>0,\delta_{2^{2n}}^j\in\mathcal{N}\}.
\end{eqnarray}
$\mathbf{J}_{\mathcal{S}_0}^\top\mathbf{Q}\delta_{2^{2n}}^j>0$ shows that there exists at least one switching signal such that $\delta_{2^{2n}}^j$ can reach $\mathcal{S}_0$. Thus, $\mathcal{N}_1$ contains all the states in $\mathcal{N}$ that can reach $\mathcal{S}_0$ in one step under at least one switching signal.

For set $\mathcal{N}$, we can find some states with the following characteristics:
each state can reach itself in one step under at least one switching signal. The definition of such a state is given below.
\begin{definition}\cite{FPIO}\label{3.1111}
A state $z_0\in\Delta_{2^{2n}}$ is called a positive-probability fixed point of PBN (\ref{8}), if $\mathbb{P}\{z(t+1)=z_0\mid z(t)=z_0\}>0, \forall t\in \mathbf{N}$.
\end{definition}

For state $\delta_{2^{2n}}^j\in\mathcal{N}$, if
\begin{align}\label{3.114}
\mathbf{Q}_{jj}>0,
\end{align}
then $\delta_{2^{2n}}^j$ is a positive-probability fixed point in set $\mathcal{N}$. Based on formula (\ref{3.114}), we can easily calculate all the positive-probability fixed points in set $\mathcal{N}$ and denoted it as
\begin{align}\label{3.115}
\mathcal{P}=\{\delta_{2^{2n}}^j\in\mathcal{N}\mid \mathbf{Q}_{jj}>0\}.
\end{align}

\begin{lemma}\label{l1.0}
To make PBN (\ref{3}) be observable, $\mathcal{N}_1\cup\mathcal{P}$ must be $H$-distinguishable.
\end{lemma}

The proof of Lemma \ref{l1.0} and all other proofs can be found in Appendix.

Let $\Gamma=\mathcal{N}_1\cup\mathcal{P}\cup\mathcal{S}_2$. To make PBN (\ref{3}) observable, $\Gamma$ should become $H$-distinguishable by adding new  measurements. According to Algorithm 1, we can calculate the reachable set of $\Gamma$ with probability one and denote it as $R(\Gamma)$. Obviously, $\mathcal{M}\subseteq R(\Gamma)$. Since $\mathcal{S}_2$ is an $H$-distinguishable set, if $\mathcal{N}_1\cup\mathcal{P}$ is also $H$-distinguishable, then $R(\Gamma)$ must be distinguishable.

Furtherly, if $\mathcal{N}\setminus (\mathcal{N}_1\cup\mathcal{P})\subseteq R(\Gamma)$, then PBN (\ref{1}) can become observable. Otherwise, let
\begin{eqnarray}
\mathcal{N}'=\mathcal{N}\setminus[\mathcal{N}_1\cup\mathcal{P}\cup R(\Gamma)],
\end{eqnarray}
then $\mathcal{N}'$ is still an indistinguishable set. We need to add appropriate measurements to make $\mathcal{N}'$ distinguishable.

However, $\mathcal{N}'$ only needs to be a distinguishable set. To make $\mathcal{N}'$ distinguishable, we do not need to make all the states in it $H$-distinguishable. We hope to find a subset $\Theta\subseteq\mathcal{N}'$ such that
\begin{align}\label{3.116}
\mathcal{N}'\backslash\Theta\subseteq R(\Theta\cup\Gamma),
\end{align}
where $R(\Theta\cup\Gamma)$ denotes the reachable set of $\Theta\cup\Gamma$ with probability one. That is, we need to find a subset $\Theta$ from $\mathcal{N}'$ such that the other states of $\mathcal{N}'$ will reach $\Theta\cup\Gamma$ with probability one. If $\Theta$ is distinguishable, then $\mathcal{N}'$ must be distinguishable. So we only need to make such a set $\Theta$ $H$-distinguishable.

It is a difficult task to find all the sets $\Theta$ that satisfy formula (\ref{3.116}). We can directly find all the nonempty subsets of $\mathcal{N}'$, and then detect which subset can satisfy formula (\ref{3.116}) one by one. Suppose that $\mathcal{N}'$ has $\gamma$ states and denote all its nonempty subsets by $\Theta_1,\Theta_2,\ldots,\Theta_{2^\gamma-1}$. According to Algorithm 1 and formula (\ref{3.116}), we can find all sets $\Theta_i, i\in[1:2^\gamma-1]$ which make $\mathcal{N}'\backslash\Theta_i$ reach $\Theta_i\cup\Gamma$ with probability one. However, the computational complexity of this method is very high, we need to determine whether $2^\gamma-1$ sets satisfy formula (\ref{3.116}). To reduce the computational complexity, the first thing is to search the states that must become $H$-distinguishable from $\mathcal{N}'$.

If $\mathcal{N}'$ has an invariant set, we can easily obtain that to make $\mathcal{N}'$ distinguishable, its invariant set must be distinguishable.
\begin{definition}\cite{SMCB}
A nonempty set $\mathcal{C}\subseteq\Delta_{2^{2n}}$ is called an invariant set with probability one of PBN (\ref{8}), if for any $z(t)\in\mathcal{C}$, it holds that
\begin{eqnarray}
\mathbb{P}\{z(t+1)\in\mathcal{C}\mid z(t)\in\mathcal{C}\}=1.
\end{eqnarray}
\end{definition}

The union of all invariant sets contained in a given set $\mathcal{C}\subseteq\Delta_{2^{2n}}$ is called
the maximum invariant set of $\mathcal{C}$, denoted by $I(\mathcal{C})$.

There have been many excellent results on the calculation of the maximum invariant set with probability one in \cite{AYCD}, \cite{AFSS}, \cite{ROTD}. Referring to these results, we can calculate the maximum invariant set $I(\mathcal{N}')$ of $\mathcal{N}'$.

If we want to make $I(\mathcal{N}')$ distinguishable,  there is no need to make all the states of $I(\mathcal{N}')$ $H$-distinguishable. We can first calculate the state set that needs to become $H$-distinguishable in $I(\mathcal{N}')$.

Since $I(\mathcal{N}')$ is an invariant set, there must be a subset in $I(\mathcal{N}')$ such that the other states in $I(\mathcal{N}')$ will reach the subset with probability one. Suppose that there are $\alpha$ states in $I(\mathcal{N}')$, then $I(\mathcal{N}')$ has $2^\alpha-1$ nonempty subsets and denote them as $\mathcal{G}_1, \mathcal{G}_2, \ldots, \mathcal{G}_{2^\alpha-1}$.
According to Algorithm 1, the reachable set of each set $\mathcal{G}_i, i=[1:2^\alpha-1]$ with probability one can be calculated respectively. Then the state set that can reach $\mathcal{G}_i$ with probability one is $R(\mathcal{G}_i)$.
If
\begin{align}
I(\mathcal{N}')\setminus\mathcal{G}_i\subseteq R(\mathcal{G}_i),
\end{align}
then $\mathcal{G}_i$ is a subset to make $I(\mathcal{N}')\backslash\mathcal{G}_i$ reach $\mathcal{G}_i$ with probability one. We denote the complete set of all sets satisfying (3.7) as $\Omega$.

In the following, Algorithm 2 is given to find set $\Omega$, where $\Omega$ contains all sets $\mathcal{G}_i$ in $I(\mathcal{N}')$ which make $I(\mathcal{N}')\backslash\mathcal{G}_i$ reach $\mathcal{G}_i$ with probability one.
\begin{algorithm}[htb]
	\renewcommand{\algorithmicrequire}{\textbf{Require:}}
	\renewcommand{\algorithmicensure}{\textbf{Ensure:}}
	\caption{:The solution method of $\Omega$ in $I(\mathcal{N}')$.}
	\label{ppp2}
	\begin{algorithmic}[1]
		\Require $I(\mathcal{N}')$
\Ensure $\Omega$
\State Calculate sets $\mathcal{G}_1, \mathcal{G}_2, \ldots, \mathcal{G}_{2^\alpha-1}$;
\State Calculate set $R(\mathcal{G}_i), i=[1:2^\alpha-1]$;
\State{\textbf{Initialize:} $\Omega=\emptyset$.}
\State {\textbf{for}
{$i=[1:2^\alpha-1]$}
\textbf{do}}
\State{\textbf{if}
$I(\mathcal{N}')\setminus\mathcal{G}_i\subseteq R(\mathcal{G}_i)$
\textbf{then}}
        \State $\Omega=\{\Omega, \mathcal{G}_i\}$;
\State \textbf{end if}
\State \textbf{end for}
	\end{algorithmic}
\end{algorithm}

We continue to observe the relationship among the sets in $\Omega$. For any sets $\mathcal{G}_i, \mathcal{G}_j\in\Omega$, if $\mathcal{G}_i\subseteq\mathcal{G}_j$, we will delete $\mathcal{G}_j$ from set $\Omega$. After the above operation, assume that $\Omega$ remains $s$ sets that are not included in each other, denoted by $\Omega_1,\Omega_2,\ldots,\Omega_s$. Then all states in $I(\mathcal{N}')\setminus\Omega_v$ can reach $\Omega_v, v\in[1:s]$ with probability one. If any a set of $\Omega_1,\Omega_2,\ldots,\Omega_s$ is $H$-distinguishable, then $I(\mathcal{N}')$ must be distinguishable.

When $I(\mathcal{N}')$ is distinguishable, we determine whether formula $\mathcal{N}'\setminus I(\mathcal{N}')\subseteq R(\Gamma\cup I(\mathcal{N}'))$ is valid. If the formula holds, then PBN (\ref{1}) is observable. Otherwise, let
\begin{align}
\mathcal{N}''=\mathcal{N}'\setminus R(\Gamma\cup I(\mathcal{N}')).
\end{align}
$\mathcal{N}''$ contains the states that are still indistinguishable in $\mathcal{N}'$ even after $\mathcal{N}_1\cup\mathcal{P}$ and $I(\mathcal{N}')$ all become distinguishable. We continue to search the states that need to become $H$-distinguishable so that $\mathcal{N}''$ becomes distinguishable.

Suppose that there are $\beta$ states in $\mathcal{N}''$, then $\mathcal{N}''$ has $2^\beta-1$ nonempty subsets and denote them as $\mathcal{\overline{G}}_1, \mathcal{\overline{G}}_2, \ldots, \mathcal{\overline{G}}_{2^\beta-1}$.
Based on Algorithm 1, we can calculate a set $\Omega'$, where $\Omega'$ contains all sets $\mathcal{\overline{G}}_i$ in $\mathcal{N}''$ which make $\mathcal{N}''\setminus\mathcal{\overline{G}}_i$ reach $\Gamma\cup R(\Gamma\cup I(\mathcal{N}'))\cup\mathcal{\overline{G}}_i$ with probability one, $i\in[0:2^\beta-1]$. For any sets $\mathcal{\overline{G}}_i, \mathcal{\overline{G}}_j\in\Omega$, if $\mathcal{\overline{G}}_i\subseteq\mathcal{\overline{G}}_j$, we will delete $\mathcal{\overline{G}}_j$ from set $\Omega'$. Assume that $\Omega'$ remains $c$ sets that are not included in each other and denote them by $\Omega'_1,\Omega'_2,\ldots,\Omega'_c$. If any a set of $\Omega'_1,\Omega'_2,\ldots,\Omega'_c$ is $H$-distinguishable, then $\mathcal{N}''$ must be distinguishable. Define
\begin{align}\label{3.113}
\Theta_v^w=\Omega_v\cup\Omega'_w,
\end{align}
where $v\in[1:s], w\in[1:c]$. It is easy to see that set $\Theta_v^w$ satisfies condition (\ref{3.116}), that is, all states in $\mathcal{N}'\backslash\Theta_v^w$ can reach $\Theta_v^w\cup\Gamma$ with probability one. If set $\Theta_v^w$ is $H$-distinguishable, then $\mathcal{N}'$ must be distinguishable, $v\in[1:s],w\in[1:c]$. Therefore, $\Theta_v^w$ is the set that we want.
\begin{theorem}\label{l2.0}
PBN (\ref{3}) becomes observable, if and only if at least one set $\mathcal{N}_1\cup\mathcal{P}\cup\Theta_v^w$ becomes $H$-distinguishable, $v\in[1:s], w\in[1:c]$.
\end{theorem}

Based on the above analysis, we have the following key steps to determine all the feasible minimal state sets that need to be $H$-distinguishable.
\begin{enumerate}
\item[$\bullet$] Step 1: Calculate the state set $\mathcal{M}$ in set $\mathcal{S}_1$ that can reach $\mathcal{S}_2$ with probability one, and obtain the indistinguishable state set $\mathcal{N}=\mathcal{S}_1\setminus \mathcal{M}$.
\item[$\bullet$]  Step 2: Identify the state set $\mathcal{N}_1\cup\mathcal{P}$ in $\mathcal{N}$ which clearly needs to be $H$-distinguishable, where $\mathcal{N}_1$ is the set that can reach $\mathcal{S}_0$ in one step under at least one switching
signal and $\mathcal{P}$ is the positive-probability fixed point set. Determine the state set $\mathcal{N}'$ which remains indistinguishable after $\mathcal{N}_1\cup\mathcal{P}$ becomes $H$-distinguishable. If $\mathcal{N}'=\emptyset$, PBN (\ref{3}) becomes observable; otherwise, go to Step 3.
\item[$\bullet$] Step 3: Calculate the maximum invariant set $I(\mathcal{N}')$ of $\mathcal{N}'$, and search the minimal state sets $\Omega_v$, $v\in[1:s]$ in $I(\mathcal{N}')$ that need to be $H$-distinguishable. Obtain the state set $\mathcal{N}''$ which remains indistinguishable after $\mathcal{N}_1\cup\mathcal{P}\cup\Omega_v$ becomes $H$-distinguishable. If $\mathcal{N}''=\emptyset$, PBN (\ref{3}) becomes observable; otherwise, go to Step 4.
    \item[$\bullet$] Step 4: Determine the minimal state subsets $\Omega'_w$, $w\in[1:c]$ in $\mathcal{N}''$ that need to be $H$-distinguishable until $\mathcal{N}''\setminus\Omega'_w$ can reach the new $H$-distinguishable set with probability one. All the minimal state sets which need to be $H$-distinguishable are $\mathcal{N}_1\cup\mathcal{P}\cup\Omega_v\cup\Omega'_w$, $v\in[1:s], w\in[1:c]$.
\end{enumerate}

\begin{remark}
According to the above process to search set $\Theta_v^w$, we only need to determine whether $2^\alpha+2^\beta-2$ sets satisfy formula (\ref{3.116}), where $\alpha+\beta\leq\gamma$. The computational complexity of this method is much lower than that of directly finding all the $2^\gamma-1$ nonempty subsets in $\mathcal{N}'$ to determine whether they satisfy formula (\ref{3.116}).
\end{remark}

\begin{remark}
In previous studies, the states that need to be $H$-distinguishable are determined by the attractors of BNs. However, since the state trajectories of a PBN from an initial state are not unique, the previous methods can not be applied to our models any more. It is much more difficult to determine the minimal sets that need to be $H$-distinguishable of a PBN than that of a BN.
\end{remark}

\subsection{The minimum observability of PBNs}
In this subsection, we consider to add the minimal measurements to make an unobservable PBN become observable.

Firstly, we discuss how to add the minimal number of measurements to make all states in a set $\mathcal{N}_1\cup\mathcal{P}\cup\Theta_v^w$ $H$-distinguishable, $v\in[1:s], w\in[1:c]$.

We first discuss which states can become $H$-distinguishable when the new measurement $y(t)=x_m(t)$ is added. Using STP, $y(t)=x_m(t)$ can be represented by algebraic form  $y(t)=H_mx(t)$, where $H_m=(\textbf{1}_{2^{m-1}}^\top\otimes I_2)(I_{2^m}\otimes\textbf{1}_{2^{n-m}}^\top)$. For state $z=\delta_{2^n}^i\delta_{2^n}^j$, if
\begin{align}\label{3.110}
H_m\delta_{2^n}^i\neq H_m\delta_{2^n}^j,
\end{align}
then state $z$ becomes $H$-distinguishable under the new measurement $y(t)=x_m(t)$. According to formula (\ref{3.110}), we can calculate a set $\Upsilon_m$ whose states are $H$-distinguishable under the new measurement $y(t)=x_m(t)$, where $\Upsilon_m=\{\delta_{2^n}^i\delta_{2^n}^j\mid H_m\delta_{2^n}^i\neq H_m\delta_{2^n}^j\}$.

Based on the above calculation process, we can find the state sets that can be $H$-distinguishable under the new output $y=x_i,i\in[1:n]$ respectively, and denoted them as $\Upsilon_1,\Upsilon_2,\ldots,\Upsilon_n$.

Choose a set $\mathcal{N}_1\cup\mathcal{P}\cup\Theta_v^w$ arbitrarily, $v\in[1:s],w\in[1:c]$. For state $z\in\mathcal{N}_1\cup\mathcal{P}\cup\Theta_v^w$, if $z\in\Upsilon_i, i\in[1:n]$, then $z$ is $H$-distinguishable under new added output $y=x_i$.
For any given state, we can find all the measurements that make it $H$-distinguishable. Assume that $\mathcal{N}_1\cup\mathcal{P}\cup\Theta_v^w=\{z_1,z_2,\ldots,z_l\}$, we can define a Boolean vector $\xi_j\in\mathcal{B}_{n\times1}$ as follows:
\begin{equation}\label{3.12}
(\xi_j)_i=
\begin{cases}
1,\  ~\mathrm{if}~ z_j\in\Upsilon_i,\\
0,\  ~\mathrm{if}~ z_j\notin\Upsilon_i.
\end{cases}
\end{equation}
Construct a truth matrix $\Phi_v^w=[\xi_1,\xi_2,\ldots,\xi_l], v\in[1:s],w\in[1:c]$. We can easily see that $(\Phi_v^w)_{i,j}=1$ if and only if $z_j\in\mathcal{N}_1\cup\mathcal{P}\cup\Theta_v^w$ is distinguishable under output $y=x_i$.
\begin{theorem}\label{3.9}
$x_{j_1},x_{j_2},\ldots,x_{j_r}$ are the minimum measurements that make all states in $\mathcal{N}_1\cup\mathcal{P}\cup\Theta_v^w$ $H$-distinguishable, if and only if $\mathcal{I}_v^w=\{j_1,j_2,\ldots,j_r\}$ is the set with the least number of elements in $\{1,2,\ldots,n\}$ such that
\begin{align}\label{3.211}
\bigvee_{i=1}^{r}\mathrm{Row}_{j_i}(\Phi_v^w)=\mathbf{1}_l^\top,
\end{align}
where $l$ is the number of elements of set $\mathcal{N}_1\cup\mathcal{P}\cup\Theta_v^w$.
\end{theorem}

\begin{remark}
Theorem \ref{3.9} provides a method for finding the least measurements that make a given set $\mathcal{N}_1\cup\mathcal{P}\cup\Theta_v^w, v\in[1:s],w\in[1:c]$ $H$-distinguishable. However, the selection of $\Theta_v^w$ is random, so these measurements are not necessarily the optimal solution of the minimum observation problem for PBN (\ref{1}).
\end{remark}

Next we will discuss how to add the least amount of measurements to make PBN (\ref{1}) observable.
For a set $\mathcal{N}_1\cup\mathcal{P}\cup\Theta_v^w, v\in[1:s],w\in[1:c]$, we can calculate all measurement sets which has the least number of elements by Theorem \ref{3.9} and denote them as $(\mathcal{I}_v^w)_1, (\mathcal{I}_v^w)_2,\ldots,(\mathcal{I}_v^w)_\beta$. Obviously,
\begin{align}\nonumber
|(\mathcal{I}_v^w)_1|=|(\mathcal{I}_v^w)_2|=\ldots=|(\mathcal{I}_v^w)_\beta|.
\end{align}
Denote the cardinality of these sets as $\lambda_v^w$.

For all $v\in[1:s],w\in[1:c]$, we can compute the cardinalities of the minimum number of measurements that make set $\mathcal{N}_1\cup\mathcal{P}\cup\Theta_v^w$ $H$-distinguishable, and denoted them as
\begin{align}\nonumber
\lambda_1^1,\lambda_1^2,\ldots,\lambda_1^c,\lambda_2^1,\lambda_2^2,\ldots,\lambda_2^c,\ldots,\lambda_s^1,\lambda_s^2,\ldots,\lambda_s^c.
\end{align}

If
\begin{align}\label{3.117}
(v^\ast,w^\ast)=argmin_{v\in[1:s],w\in[1:c]}\lambda_v^w,
\end{align}
then as long as we make set $\mathcal{N}_1\cup\mathcal{P}\cup\Theta_{v^\ast}^{w^\ast}$ $H$-distinguishable, it can be guaranteed that the added measurements are the minimum. Suppose that $\mathcal{I}_{v^\ast}^{w^\ast}=\{j_1^\ast,j_2^\ast,\ldots,j_r^\ast\}$ is the set with the least number of elements in $\{1,2,\ldots,n\}$ such that all states in $\mathcal{N}_1\cup\mathcal{P}\cup\Theta_{v^\ast}^{w^\ast}$ are $H$-distinguishable,
then $x_{j_1^\ast},x_{j_2^\ast},\ldots,x_{j_r^\ast}$ are the minimum measurements that make the whole system (\ref{1}) observable.

\begin{remark}
It is worth mentioning that the minimum measurements of a PBN are not necessarily unique,
we can choose the appropriate measurements to make the system observable according to the actual demand and observation cost.
\end{remark}
\begin{remark}
This paper takes into account all the feasible schemes for adding measurements required to make the system observable, and then obtains the minimum measurements needed by comparison. Our results provide more feasible schemes for observer design than previous ones.
\end{remark}

\section{An illustrative example}

{\example Consider the apoptosis network modeled by PBNs in \cite{SEFP}, which consists of the following four subnetworks:
\begin{eqnarray}\label{4.11}
\begin{cases}
x_1(t+1)=f_1^\sigma(x_1(t),x_2(t),x_3(t)),&\\
x_2(t+1)=f_2^\sigma(x_1(t),x_2(t),x_3(t)),&\\
x_3(t+1)=f_3^\sigma(x_1(t),x_2(t),x_3(t)),&\\
\end{cases}
\end{eqnarray}
where $x_1\in\mathcal{D}$ denotes the concentration level of apoptosis
proteins, $x_2\in\mathcal{D}$ denotes state level presenting the active level of caspase 3, $x_3\in\mathcal{D}$ presents the active level of caspase 8.
The switching signal $\sigma$ is a stochastic sequence taking values from [1 : 4], and

$\begin{cases}
f_1^1=\neg x_2(t),\\
f_2^1=\neg x_1(t)\wedge\neg x_3(t),\\
f_3^1=x_2(t)),\ \ \ \ \ \ \ \ \ \ \ \
\end{cases}$
$\begin{cases}
f_1^2=\neg x_2(t),\\
f_2^2=x_2(t),\\
f_3^2=x_2(t),\ \ \ \ \ \ \ \ \
\end{cases}$
$\begin{cases}
f_1^3=x_1(t),\\
f_2^3=\neg x_1(t)\wedge\neg x_3(t),\\
f_3^3=x_2(t),
\end{cases}$\ \ \ \ \
$\begin{cases}
f_1^4=x_1(t),\\
f_2^4=x_2(t),\\
f_3^4=x_2(t).\\
\end{cases}$
\ \\
The probability distribution of each subnetwork is
$\mathbb{P}\{f=f^1\}=0.27,\ \mathbb{P}\{f=f^2\}=0.03,\ \mathbb{P}\{f=f^3\}=0.63, \ \mathbb{P}\{f=f^4\}=0.07$, where $f^i=\{f_1^i,f_2^i,f_3^i\}$, $i\in[1:4].$}

The output of system (\ref{4.11}) is as follows:
\begin{eqnarray}\nonumber
y(t)=(x_2\wedge\neg x_3)\vee(x_1\leftrightarrow x_3\wedge\neg x_2).
\end{eqnarray}
Identify $1\sim\delta_2^1, 0\sim\delta_2^2.$ Then, the algebraic form of (\ref{4.11}) can be expressed as
\begin{eqnarray}\label{4.1}
\begin{cases}
x(t+1)=\mathbf{L}_{\sigma(t)}x(t),\\
y(t)=Hx(t),
\end{cases}
\end{eqnarray}
where $x(t)=\ltimes_{i=1}^3x_i(t)\in\Delta_8$, $y(t)\in\Delta_2$, and $\mathbf{L}_i\in\mathcal{L}_{8\times8}, i\in[1:4]$ is the structure matrix of the $i$-th subnetwork, $H\in\mathcal{L}_{2\times8}$, which are shown as
\begin{align}\nonumber
&\mathbf{L}_1=\delta_8[7\ 7\ 4\ 4\ 7\ 5\ 4\ 2],\ \ \ \mathbf{L}_2=\delta_8[5\ 5\ 4\ 4\ 5\ 5\ 4\ 4],\\
&\mathbf{L}_3=\delta_8[3\ 3\ 4\ 4\ 7\ 5\ 8\ 6],\ \ \ \mathbf{L}_4=\delta_8[1\ 1\ 4\ 4\ 5\ 5\ 8\ 8],\\ \nonumber
&H=\delta_2[2\ 1\ 1\ 2\ 2\ 1\ 2\ 1]. \nonumber
\end{align}
According to the probability distribution of each subnetwork, we can get the probability distribution vector of (\ref{4.11}) as $\mathbf{p}^\sigma=[0.27\ 0.03\ 0.63\ 0.07]^\top$.

Let $\mathbf{L}=[\mathbf{L}_1\ \mathbf{L}_2\ \mathbf{L}_3\ \mathbf{L}_4]$. The expectation of the augmented system can be calculated as
\begin{eqnarray}\nonumber
\mathbb{E}\{z(t+1)\}=\mathbf{Q}\mathbb{E}\{z(t)\},
\end{eqnarray}
where $\mathbf{Q}=\{[\mathbf{L}(I_{4\times2^3}\otimes\mathbf{1}_{2^3}^\top)]\ast[\mathbf{L}(I_4\otimes\mathbf{1}_{2^3}^\top)]\}
\mathbf{p}^\sigma$, and the columns of $\mathbf{Q}$ are shown in Table 2.

\begin{table}[H]
\centering
\caption{Transition probability matrix of (\ref{4.1})}
\vspace{2.5mm}
\scriptsize
\begin{tabular}{cc}
\hline
Column index $j$ & $j$th column of $\mathbf{Q}$\\
\hline
$j\in\mathcal{S}_1$ & $\backslash$ \\
 $j=4$ &  $0.07\delta_{64}^4+0.63\delta_{64}^{20}+0.03\delta_{64}^{36}++0.27\delta_{64}^{52}$\\
$j=5$ &  $0.07\delta_{64}^5+0.63\delta_{64}^{23}+0.03\delta_{64}^{37}++0.27\delta_{64}^{55}$\\
$j=7$ &  $0.07\delta_{64}^8+0.63\delta_{64}^{24}+0.03\delta_{64}^{36}++0.27\delta_{64}^{52}$\\
$j=11$ &  $0.07\delta_{64}^4+0.63\delta_{64}^{20}+0.03\delta_{64}^{36}++0.27\delta_{64}^{52}$\\
$j=14$ & $0.07\delta_{64}^5+0.63\delta_{64}^{21}+0.03\delta_{64}^{37}++0.27\delta_{64}^{53}$\\
$j=16$ & $0.07\delta_{64}^8+0.63\delta_{64}^{22}+0.03\delta_{64}^{36}++0.27\delta_{64}^{50}$\\
$j=22$ &  $\delta_{64}^{29}$\\
$j=24$ & $0.27\delta_{64}^{26}+0.03\delta_{64}^{28}+0.63\delta_{64}^{30}++0.07\delta_{64}^{32}$\\
$j=29$ & $0.1\delta_{64}^{29}+0.9\delta_{64}^{31}$\\
$j=31$ & $0.3\delta_{64}^{28}+0.7\delta_{64}^{32}$\\
$j=48$ &  $0.27\delta_{64}^{34}+0.03\delta_{64}^{36}+0.63\delta_{64}^{38}++0.07\delta_{64}^{40}$\\
\hline
\end{tabular}
\end{table}

\begin{figure}
  \begin{center}
  \includegraphics[width=8.0 cm,height=4.2 cm, bb=0 70 550 375]{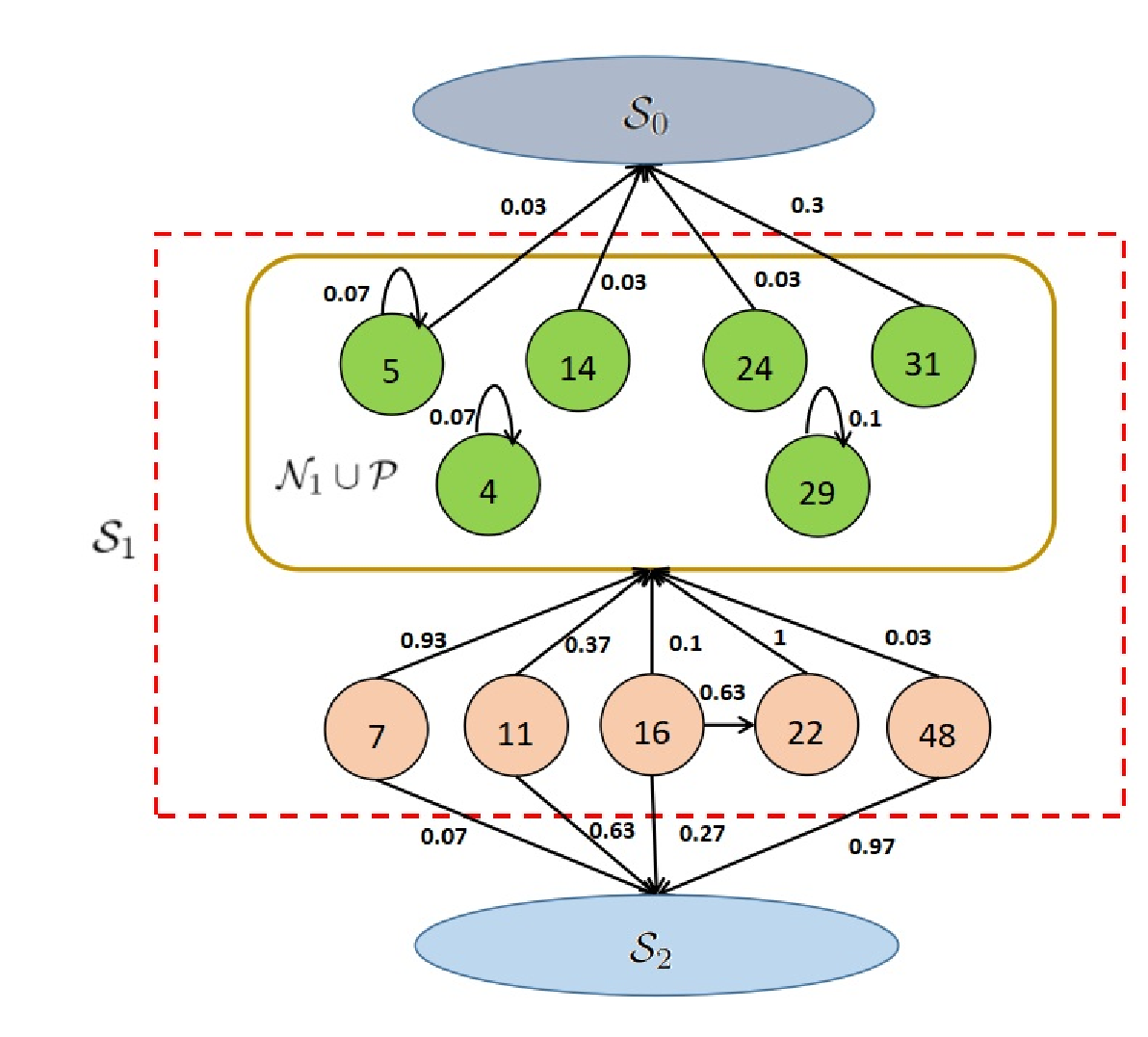}\\
  \vskip -4mm
  \vspace{\baselineskip}
  \vspace{\baselineskip}
 \caption{The state transfer graph of $\mathcal{S}_1$.}
 \label{fig1}
 \end{center}
\end{figure}

According to (\ref{9}), it is easy to obtain that
\begin{align}\nonumber
&\mathcal{S}_0=\{\delta_{64}^1,\delta_{64}^{10},\delta_{64}^{19},\delta_{64}^{28},\delta_{64}^{37},\delta_{64}^{46},\delta_{64}^{55},\delta_{64}^{64}\},&\\ \nonumber
&\mathcal{S}_1=\{\delta_{64}^{4},\delta_{64}^{5},\delta_{64}^{7},\delta_{64}^{11},\delta_{64}^{14},\delta_{64}^{16},\delta_{64}^{22},\delta_{64}^{24},\delta_{64}^{29},\delta_{64}^{31},\delta_{64}^{48}\},&\\ \nonumber
&\mathcal{S}_2=\{\delta_{64}^{2},\delta_{64}^{3},\delta_{64}^{6},\delta_{64}^{8},\delta_{64}^{12},\delta_{64}^{13},\delta_{64}^{15},\delta_{64}^{20},\delta_{64}^{21},\delta_{64}^{23},\delta_{64}^{30},\delta_{64}^{32},\delta_{64}^{38},\delta_{64}^{40},\delta_{64}^{47},\delta_{64}^{56}\}.&\nonumber
\end{align}
The state evolutionary trajectories of $\mathcal{S}_1$ can be shown in Figure \ref{fig1}.

From formulas (\ref{3.13}) and (\ref{3.115}), we can know that $\mathcal{N}_1=\{\delta_{64}^{5},\delta_{64}^{14},\delta_{64}^{24},\delta_{64}^{31}\}$ and
$\mathcal{P}=\{\delta_{64}^{4},\delta_{64}^{5},\delta_{64}^{29}\}$, then
\begin{eqnarray}\nonumber
\mathcal{N}_1\cup\mathcal{P}=\{\delta_{64}^{4},\delta_{64}^{5},\delta_{64}^{14},\delta_{64}^{24},\delta_{64}^{29},\delta_{64}^{31}\}.
\end{eqnarray}
It is known from Lemma \ref{l1.0} that to make PBN (\ref{4.11}) observable, $\mathcal{N}_1\cup\mathcal{P}$ must be $H$-distinguishable. Since $\delta_{64}^{13}$ and $\delta_{64}^{34}$, $\delta_{64}^{15}$ and $\delta_{64}^{50}$, $\delta_{64}^{29}$ and $\delta_{64}^{36}$, $\delta_{64}^{31}$ and $\delta_{64}^{52}$ have the same distinguishability, respectively, we just consider states $\delta_{64}^{13}$, $\delta_{64}^{15}$, $\delta_{64}^{29}$ and $\delta_{64}^{31}$. If $\mathcal{N}_1\cup\mathcal{P}$ becomes $H$-distinguishable, it can be seen that $\delta_{64}^7, \delta_{64}^{11}, \delta_{64}^{16}, \delta_{64}^{22}, \delta_{64}^{48}$ can reach the $H$-distinguishable states with probability one by Figure \ref{fig1}, then all states in $\mathcal{S}_1$ can be distinguishable. Thus, as long as all states in $\mathcal{N}_1\cup\mathcal{P}$ become $H$-distinguishable, then PBN (\ref{4.11}) is observable.

Next, we make $\mathcal{N}_1\cup\mathcal{P}$ $H$-distinguishable by adding new measurements. We take the first state $\delta_{64}^4$ as an example. Since $\delta_{64}^4$ can be decomposed into $\delta_2^1\delta_2^1\delta_2^1$ and $\delta_2^1\delta_2^2\delta_2^2$, then we know that $\delta_{64}^4$ becomes $H$-distinguishable under new measurements $y=x_2$ and $y=x_3$ based on formula (\ref{3.110}) and (\ref{3.12}), $\delta_{64}^4\in\Upsilon_i, i\in\{2,3\}.$ So $\xi_1=[0\ 1\ 1]^\top$. Similarly, we can get that $\xi_2=[1\ 0\ 0]^\top$, $\xi_3=[1\ 0\ 0]^\top$, $\xi_4=[1\ 0\ 1]^\top$, $\xi_5=[1\ 1\ 1]^\top$, $\xi_6=[1\ 0\ 0]^\top$. It can be obtained that
\begin{eqnarray}\nonumber
\Phi=\left[
  \begin{array}{ccccccccccc}
  0\ & 1\ & 1\ & 1\ & 1\ & 1\ \\
  1\ & 0\ & 0\ & 0\ & 1\ & 0\ \\
  1\ & 0\ & 0\ & 1\ & 1\ & 0\ \\
  \end{array}
\right].
\end{eqnarray}
By Theorem \ref{3.9}, we can easily see that
\begin{eqnarray}\nonumber
\mathrm{Row}_1(\Phi)\vee\mathrm{Row}_i(\Phi)=\mathbf{1}_6^\top, i\in\{2,3\}.
\end{eqnarray}
Therefore, $x_1,x_2$ or $x_1,x_3$ are the minimum measurements that make PBN (\ref{4.1}) observable. We can consider the cost of caspase 3 and caspase 8 to add appropriate measurements to make system (\ref{4.1}) observable.

\section{Conclusion}
In this study, we have investigated the minimum observability of PBNs. Applying the STP method, we have established an augmented system and converted the observability problem of the original PBN  into the set reachability problem of the augmented system. Then the minimum observability of the PBN has been solved by two main steps: determine the states that need to be $H$-distinguishable from the  augmented system and add the least measurements to make a give state set $H$-distinguishable. For the first issue, we have found all the minimal $H$-distinguishable state sets which can make the originally unobservable PBN become observable. And for the second issue, we have proposed a necessary and sufficient condition for finding the least measurements to make a given set $H$-distinguishable. Furthermore, by comparing the numbers of measurements
for all the feasible $H$-distinguishable state sets, we have obtained a method to determine all the minimum measurements to make the system observable. Compared with the existing minimum observable results, our method provide more feasible schemes for observer design.

We mainly apply the robust reachable set technique to find the minimum measurements. Although we greatly reduce the computational complexity by means of positive probability fixed points and invariant sets, it is still not a simple calculation process. In future research, we will consider the reducibility of  measurements so that the computational complexity can be reduced further.

\section{Acknowledgements}
This work was supported by the National Natural Science Foundation of China under Grant 62103176, the ``Guangyue Young Scholar Innovation Team" of Liaocheng University under Grant LCUGYTD2022-01, and the ``Discipline with Strong Characteristics of Liaocheng University-Intelligent Science and Technology" under Grant 319462208.

\renewcommand\baselinestretch{1.3}

\section{Appendix}

\emph{Proof of Lemma \ref{l1.0}:}
We first prove that $\mathcal{N}_1$ must become $H$-distinguishable. Assume that there exists a state $z_0\in\mathcal{N}_1$ which is not $H$-distinguishable. Since $z_0\in\mathcal{N}_1$, there is a switching signal $\sigma^1\in\Sigma^1$ such that $z(1;\sigma^1,z_0)\in\mathcal{S}_0$. Obviously, for any integer $t\geq1$, $z(t;\sigma^t,z_0)\in\mathcal{S}_0$ holds. Suppose that $z_0=x_0\ltimes x'_0$, then $x(t;\sigma^t,x_0)\ltimes x'(t;\sigma^t,x'_0)\in\mathcal{S}_0$ for any $t\in\mathbf{Z}_+$, which shows that $y(t;\sigma^t,x_0)=y'(t;\sigma^t,x'_0)$. So the PBN is unobservable. Therefore, $\mathcal{N}_1$ must be $H$-distinguishable.

Then we prove that $\mathcal{P}$ must be $H$-distinguishable. Suppose that there exists a positive-probability fixed point $z_0\in\mathcal{P}$ which is $H$-indistinguishable. According to Definition \ref{3.1111}, we can know that
\begin{align}\nonumber
\mathbb{P}\{z(t+1)=z_0\mid z(t)=z_0\}>0.
\end{align}
Hence, for any $t\in\mathbf{Z}_+$, $z_0$ can always reach itself with a positive probability. Since $z_0$ can not reach any other states with probability one, as long as $z_0$ is not $H$-distinguishable, it can not become a distinguishable state. So $\mathcal{N}$ can not become a distinguishable set. Thus, $\mathcal{P}$ must be $H$-distinguishable.
The lemma is established. $\Box$

\emph{Proof of Theorem \ref{l2.0}:}
(Sufficiency) If the states in set $\mathcal{N}_1\cup\mathcal{P}\cup\Theta_v^w$ are $H$-distinguishable, $v\in[1:s], w\in[1:c]$, then the states of $R(\Gamma)$ and $\mathcal{N}'$ are distinguishable.
Since $\mathcal{N}=\mathcal{N}_1\cup\mathcal{P}\cup (R(\Gamma)\cap\mathcal{N})\cup\mathcal{N}'$, then all states in $\mathcal{N}$ are distinguishable. Therefore, PBN (\ref{1}) is observable.

(Necessity) Suppose there is no set $\mathcal{N}_1\cup\mathcal{P}\cup\Theta_v^w$ which is $H$-indistinguishable, $v\in[1:s], w\in[1:c]$, then $\mathcal{N}_1\cup\mathcal{P}$ is not $H$-distinguishable, or $\mathcal{N}_1\cup\mathcal{P}$ is $H$-distinguishable, but all sets of $\Theta_v^w$ are H-indistinguishable. If $\mathcal{N}_1\cup\mathcal{P}$ is not $H$-distinguishable, then PBN (\ref{1}) is unobservable by Lemma \ref{l1.0}. If $\Theta_v^w$ is $H$-indistinguishable, then $\mathcal{N}'$ is indistinguishable. Thus PBN (\ref{1}) is unobservable. In summary, to make PBN (\ref{1}) becomes observable, at least one set $\mathcal{N}_1\cup\mathcal{P}\cup\Theta_v^w$ must be $H$-distinguishable.  $\Box$

\emph{Proof of Theorem \ref{3.9}:}
(Necessity) Suppose that $x_{j_1},x_{j_2},\ldots,x_{j_r}$ are the minimum measurements that make all states in $\mathcal{N}_1\cup\mathcal{P}\cup\Theta_v^w$ $H$-distinguishable, where $\mathcal{N}_1\cup\mathcal{P}\cup\Theta_v^w=\{z_1,z_2,\ldots,z_l\}$. Then for any state $z_c\in\mathcal{N}_1\cup\mathcal{P}\cup\Theta_v^w$, there is at least one element $j_k\in\{j_1,j_2,\ldots,j_r\}$, such that
\begin{align}\nonumber
(\Phi_v^w)_{j_k,c}=1.
\end{align}
Since there are no fewer measurements that make all states in $\mathcal{N}_1\cup\mathcal{P}\cup\Theta_v^w$ $H$-distinguishable, therefore, $\{j_1,j_2,\ldots,j_r\}$ is the set with the least number of elements that makes formula (\ref{3.211}) hold.

(Sufficiency) Assume that $\mathcal{I}_v^w=\{j_1,j_2,\ldots,j_r\}$ is the set with the least number of elements in $\{1,2,\ldots,n\}$ such that
$\bigvee_{i=1}^{r}\mathrm{Row}_{j_i}(\Phi_v^w)=\mathbf{1}_l^\top$. Then for any set $\{i_1,i_2,\ldots,i_k\}\subseteq\{1,2,\ldots,n\}$, where $k<r$, it holds that $\bigvee_{a=1}^{k}\mathrm{Row}_{i_a}(\Phi_v^w)\neq\mathbf{1}_l^\top$. This means that there is at least one state $z_\mu\in\mathcal{N}_1\cup\mathcal{P}\cup\Theta_v^w$ such that $(\Phi_v^w)_{i_\nu,\mu}=0$ for any $i_\nu\in\{i_1,i_2,\ldots,i_k\}$. Then state $z_\mu$ is indistinguishable under measurements $x_{i_1},x_{i_2},\ldots,x_{i_k}$. This shows that $x_{i_1},x_{i_2},\ldots,x_{i_k}$ cannot make all the states of set $\mathcal{N}_1\cup\mathcal{P}\cup\Theta_v^w$ become $H$-distinguishable. Therefore, $x_{j_1},x_{j_2},\ldots,x_{j_r}$ are the minimum measurements that make all states in $\mathcal{N}_1\cup\mathcal{P}\cup\Theta_v^w$ $H$-distinguishable.
$\Box$

The code of the probability transition matrix $\mathbf{Q}$ in the example is as follows.
\begin{itemize}
  \item[] $s1=kron(eye(2),[0\ 1;1\ 0])$;
  \item[] $s2=kron(eye(2),[1\ 1])$;
  \item[] $s3=kron([1\ 1], eye(2))$;
  \item[] $a=spn([1\ 0\ 0\ 0;0\ 1\ 1\ 1],[0\ 1;1\ 0],s1,s2)$;
  \item[] $s=kron(s3,[1\ 1])$;
  \item[] $c=kron(eye(2),[1\ 1\ 1\ 1])$;
  \item[] $L1=mkmkn(spn([0\ 1;1\ 0],s),a,s)$;
  \item[] $L2=mkmkn(spn([0\ 1;1\ 0],s),s,s)$;
  \item[] $L3=mkmkn(c,a,s)$;
  \item[] $L4=mkmkn(c,s,s)$;
  \item[] $L=[L1\ L2\ L3\ L4]$;
  \item[] $o=kron(eye(4),[1\ 1\ 1\ 1\ 1\ 1\ 1\ 1])$;
  \item[] $f=kron(o,eye(8))$;
  \item[] $g=kron(eye(32),[1\ 1\ 1\ 1\ 1\ 1\ 1\ 1])$;
  \item[] $w1=spn(L,g)$;
  \item[] $w2=spn(L,f)$;
  \item[] $Q1=mkmkn(w1,w2)$;
  \item[] $lm(Q1)$;
  \item[] $Q=spn(Q1,[0.27;0.03;0.63;0.07])$;
\end{itemize}

All the codes can be derived from the STP Toolbox.

Please refer to website \url{https://lsc.amss.ac.cn/~hsqi/soft/STP.zip} for the STP Toolbox or download it from Pro. Qi's personal homepage:
\url{http://lsc.amss.ac.cn/~hsqi/}.


\small
\begin{thebibliography}{11}


\bibitem{MSEI} K. Kauffman, ``Metabolic stability and epigenesis in randomly constructed genetic nets,'' {\it Journal of Theoretical Biology}, 1969, 22(3):  437-467.

\bibitem{HRAA} T. Akutsu, M. Hayashida, W. Ching, ``Control of Boolean networks: hardness results and algorithms for tree structured networks,'' {\it Journal of Theoretical Biology}, 2007, 244(4): 670-679.

\bibitem{OLGR} H. L\"{a}hdesm\"{a}ki, I. Shmulevich, and O. Yli-Harja, ``On learning gene regulatory networks under the Boolean network model,'' {\it Machine Learning}, 2003, 52: 147-167.

\bibitem{DRWS} A. Nazi, M. Raj, M. Francesco, P. Ghoshet and S. Das, ``Deployment of robust wireless sensor networks using gene regulatory networks: an isomorphism-based approach,'' {\it Pervasive And Mobile Computing}, 2014, 13: 246-257.

\bibitem{TEOS} D. Green, T. Leishman, S. Sadedin, ``The emergence of social consensus in Boolean networks,'' {\it IEEE Symposium on Artificial Life}, 2007, 402-408.

\bibitem{ANSM} J. Lu, B. Li, J. Zhong, ``A novel synthesis method for reliable feedback shift registers via Boolean networks,'' {\it Science China Information Sciences}, 2021, 64: 1-14.

\bibitem{AMSF} H. Zheng, D. Shi, ``A multi-agent system for environmental monitoring using boolean networks and reinforcement learning,'' {\it Journal of Cybersecurity}, 2020, 2(2): 85.

\bibitem{PBNA} I. Shmulevich, E. Dougherty, S. Kim, W. Zhang, ``Probabilistic Boolean networks: a rule-based uncertainty model for gene regulatory networks,'' {\it Bioinformatics}, 2002, 18(2): 261-274.

\bibitem{IICS} R. Pal, A. Datta, M. Bittner, ``Intervention in context-sensitive probabilistic Boolean networks,'' {\it Bioinformatics}, 2005, 21(7): 1211-1218.

\bibitem{APCA} Q. Liu, ``An optimal control approach to probabilistic Boolean networks,'' {\it Physica A-Statistical Mechanics And Its Applications}, 2012, 39(24): 6682-6689.

\bibitem{OBAT} K. Kobayashi, K. Hiraishi, ``Optimization-based approaches to control of probabilistic Boolean networks,'' {\it Algorithms}, 2017, 10(1): 31.

\bibitem{SAOG} I. Shmulevich, I. Gluhovsky, R. Hashimoto, ``Steady-state analysis of genetic regulatory networks modelled by probabilistic Boolean networks,'' {\it Comparative and functional genomics}, 2003, 4(6): 601-608.

\bibitem{PAOP} K. Kobayashi, K. Hiraishi, ``Reachability analysis of probabilistic Boolean networks using model checking,'' {\it Proceedings of SICE Annual Conference}, IEEE, 2010, 829-832.

\bibitem{ALRD} D. Cheng and H. Qi,  ``A linear representation of dynamics of Boolean networks,'' {\it IEEE Transactions On Automatic Control}, 2010, 55(10): 2251-2258.

\bibitem{C1} D. Cheng and H. Qi, ``Controllability and observability of Boolean control networks,'' {\it Automatica}, 2009, 45(7): 1659-1667.

\bibitem{C2} Y. Liu, H. Chen, J. Lu and B. Wu, ``Controllability of probabilistic Boolean control networks based on transition probability matrices",  {\it Automatica}, 2015, 52: 340-345.

\bibitem{C3} H. Li, S. Wang, X. Li and G. Zhao, ``Perturbation analysis for controllability of logical control networks,'' {\it Siam Journal On Control And Optimization}, 2020, 58(6): 3632-3657.

\bibitem{C4} S. Zhu, J. Lu, S. Azuma, and W. Zheng, ``Strong Structural Controllability of Boolean Networks: Polynomial-Time Criteria, Minimal Node Control, and Distributed Pinning Strategies,'' {\it IEEE Transactions on Automatic Control}, 2022, 3226701.

\bibitem{o1} Y. Guo, ``Observability of Boolean control networks using parallel extension and set reachability,'' {\it IEEE Transactions on Neural Networks and Learning Systems}, 2018, 29(12): 6402-6408.

\bibitem{o2} E. Fornasini and M. Valcher, ``Observability and reconstructibility of probabilistic Boolean networks,'' {\it IEEE Control Systems Letters}, 2020, 4(2): 319-324.

\bibitem{o5} S. Zhu, J. Lu, J. Zhong, Y. Liu, and J. Cao, ``On the Sensors Construction of Large Boolean Networks via Pinning Observability,'' {\it IEEE Transactions on Automatic Control}, 2022, 67(8): 4162-4169.

\bibitem{d1} B. Wang and J. Feng, ``On detectability of probabilistic Boolean networks,'' {\it Information Sciences}, 2019, 483: 383-395.

\bibitem{d2} B. Wang and J. Feng, ``Detectability of Boolean networks with disturbance inputs," {\it Systems And Control Letters},  2020, 145: 104783.

\bibitem{SSAS} Y. Guo, R. Zhou, Y. Wu, W. Gui and C. Yang, ``Stability and set stability in distribution of probabilistic Boolean networks,'' {\it IEEE Transactions on Automatic Control}, 2018, 64(2): 736-742.

\bibitem{li3} H. Li, X. Yang, S. Wang, ``Robustness for stability and stabilization of Boolean networks with stochastic function perturbations,'' {\it IEEE Transactions on Automatic Control}, 2021, 66(3): 1231-1237.

\bibitem{SFSF} R. Li, M. Yang, and T. G. Chu, ``State feedback stabilization for probabilistic Boolean networks,'' {\it Automatica}, 2014, 50(4): 1272-1278.

\bibitem{F1} J. Feng, Y. Li, S. Fu and H. Lyu, ``New method for disturbance decoupling of Boolean networks,'' {\it IEEE Transactions on Automatic Control}, 2022, 3161609.

\bibitem{wu2} Y. Wu and T. Shen, ``Policy iteration algorithm for optimal control of stochastic logical dynamical systems,'' {\it IEEE Transactions On Neural Networks And Learning Systems}, 2018, 29(5): 2031-2036.

\bibitem{li1} H. Li, X. Yang, ``Robust optimal control of logical control networks with function perturbation,'' {\it Automatica}, 2023, 152: 110970.

\bibitem{li2} S. Wang, H. Li, ``Aggregation method to reachability and optimal control of large-size Boolean control networks,'' {\it Science China Information Sciences.}, 2023, 66(7): 179202.

\bibitem{z1} G. Zhao, Y. Wang, H. Li, ``A matrix approach to the modeling and analysis of networked evolutionary games with time delays,'' {\it IEEE/CAA Journal of Automatica Sinica}, 2016, 5(4): 818-826.

\bibitem{e1} Y. Wu, S. Le, K. Zhang, X. Sun, ``Ex-ante agent transformation of Bayesian games,'' {\it IEEE Transactions on Automatic Control}, 2022, 67(11): 5793-5808.

\bibitem{S} Y. Yan, D. Cheng, J. Feng, H. Li, and J. Yue, ``Survey on applications of algebraic state space theory of logical systems to finite state machines,'' {\it Science China Information Sciences}, 2023, 66(1): 111201.

\bibitem{OOBN} D. Laschov, M. Margaliot, G. Even, ``Observability of Boolean networks: A graph-theoretic approach,'' {\it Automatica} 2013, 49(8), 2351-2362.

\bibitem{OOBM} Y. Wu, J. Xu, X. Sun, W. Wang, ``Observability of Boolean multiplex control networks,'' {\it Scientific Reports}, 2017, 7(46495): 1-15.

\bibitem{OBNV} D. Cheng, C. Li, F. He, ``Observability of Boolean networks via set
controllability approach,'' {\it Systems And Control Letters}, 2018, 115: 22-25.

\bibitem{OBUP} Y. Guo, ``Observability of Boolean control networks using parallel extension and set reachability,'' {\it IEEE Transactions on Neural Networks and Learning Systems}, 2018, 29(12): 6402-6408.

\bibitem{OOBC} K. Zhang, L. Zhang, ``Observability of Boolean control networks: A unified approach based on finite automata,'' {\it IEEE Transactions on Automatic Control} 2016, 61(9): 2733-2738.

\bibitem{EVOO} K. Zhang, K. Johansson, ``Efficient verification of observability and reconstructibility for largeBoolean control networks with special structures,'' {\it IEEE Transactions on Automatic Control}, 2020, 65(12): 5144-5158.

\bibitem{OOPB} J. Zhao, Z. Liu, ``Observability of probabilistic Boolean networks,'' {\it Chinese Control Conference}, 2015, 183-186.

\bibitem{pin5}  R. Zhou, Y. Guo, and W. Gui, ``Set reachability and observability of probabilistic Boolean networks,'' {\it Automatica}, 2019, 106: 230-241.

\bibitem{OARO} E. Fornasini, M. Valcher, ``Observability and reconstructibility of probabilistic Boolean networks,'' {\it IEEE Control Systems Letters}, 2020, 4(2): 319-324.

\bibitem{APAF} E. Weiss, M. Margaliot, ``A polynomial-time algorithm for solving the minimal observability problem in conjunctive Boolean networks,'' {\it Transactions On Automatic Control}, 2019, 64: 2727-2736.

\bibitem{o3} Y. Liu, L. Wang, Y. Yang and Z. Wu, ``Minimal observability of Boolean control networks", {\it Systems And Control Letters}, 2022, 163: 105204.

\bibitem{o4} Y. Liu, J. Zhong, ``Minimal observability of Boolean networks", {\it Science China Information Sciences} 2022, 65(5): 152203.

\bibitem{ZALJ} J. Feng, J. Yao, P. Cui, ``Singular Boolean networks: semi-tensor product approach,'' {\it Science China Information Sciences}, 2013, 56: 1-14.

\bibitem{FPIO} X. Yang, H. Li. ``Function perturbation impact on asymptotical stability of probabilistic Boolean networks: Changing to finite-time stability,'' {\it Journal of the Franklin Institute}, 2020, 357(15): 10810-10827.

\bibitem{SMCB} A. Julius, A. Halasz, M. Sakar, H. Rubin, V. Kumar and G. Pappas, ``Stochastic modeling and control of biological systems: The lactose regulation system of Escherichia coli,'' {\it IEEE Transactions On Automatic Control}, 2008, 53: 51-65.

\bibitem{AYCD} Y. Guo, R. Zhou, Y. Wu, W. Gui and C. Yang.  ``Stability and set stability in distribution of probabilistic Boolean networks,'' {\it IEEE Transactions on Automatic Control}, 2019, 64(2): 736-742


\bibitem{AFSS} R. Zhou, Y. Guo, Y. Wu, et al. ``Asymptotical feedback set stabilization of probabilistic Boolean control networks,'' {\it IEEE transactions on neural networks and learning systems}, 2019, 31(11): 4524-4537.

\bibitem{ROTD} Y. Guo, R. Zhou, Y. Wu, et al. ``Stability and set stability in distribution of probabilistic Boolean networks,'' {\it IEEE Transactions on Automatic Control}, 2018, 64(2): 736-742.

\bibitem{SEFP} J. Zhong, Z. Yu, Y. Li, et al. ``State estimation for probabilistic Boolean networks via outputs observations,'' {\it IEEE Transactions on Neural Networks and Learning Systems}, 2021, 33(9): 4699-4711.

\end{thebibliography}
\end{document}